\documentclass[iop]{emulateapj}
\usepackage{natbib}
\usepackage{color}
\usepackage[english]{babel}
\usepackage[normalem]{ulem}
\usepackage{blindtext}
\usepackage{textgreek}
\newcommand{\etal}{et\,al.}
\newcommand{\halpha}{H$\alpha$}
\newcommand{\HI}{H~{\sc I}} 
\newcommand{\lsim}{\raise0.3ex\hbox{$<$}\kern-0.75em{\lower0.65ex\hbox{$\sim$}}}
\newcommand{\gsim}{\raise0.3ex\hbox{$>$}\kern-0.75em{\lower0.65ex\hbox{$\sim$}}}
\newcommand{\msun}{M$_{\odot}$}

\newcommand{\kms}{km\,s$^{-1}$}
\begin{document}
\slugcomment{Accepted for publication in the Astrophysical Journal Letters}
%-----------------------------------------------------------------------------%
\title{Delayed Stellar Mass Assembly in the Low Surface Brightness
  Dwarf Galaxy KDG\,215}\footnote{Based on observations
  made with the NASA/ESA Hubble Space Telescope, obtained from the
  Data Archive at the Space Telescope Science Institute, which is
  operated by the Association of Universities for Research in
  Astronomy, Inc., under NASA contract NAS 5-26555.}
%-----------------------------------------------------------------------------%

%-----------
\author{John M. Cannon}
\affil{Department of Physics \& Astronomy, Macalester College, 
1600 Grand Avenue, Saint Paul, MN 55105, USA}

\author{Zili Shen, Kristen B.W. McQuinn}
\affil{University of Texas at Austin, McDonald Observatory, 2515 Speedway, Stop C1400 Austin, TX 78712, USA}

\author{Joshua Bartz, Lilly Bralts-Kelly, Alyssa M. Bulatek, Sarah
  Chinski, Robert N. Ford, Alex J.~R. Gordon, Greta Helmel, Sam Hollenbach,
  Riley A. McGlasson, Andrew Mizener, Tylyn Page, William Retza, Moritz
  Rusch, Sarah Taft}
\affil{Department of Physics \& Astronomy,
  Macalester College, 1600 Grand Avenue, Saint Paul, MN 55105, USA}

\author{Andrew E. Dolphin}
\affil{Raytheon Company, PO Box 11337, Tucson, AZ 85734, USA}

\author{Igor Karachentsev}
\affil{Special Astrophysical Observatory of RAS, Nizhnij Arkhyz, KChR, 369167, Russia}

\author{John J. Salzer}  
\affil{Department of Astronomy, Indiana University, 727 East
  Third Street, Bloomington, IN 47405, USA}

%-----------------------------------------------------------------------------%
\begin{abstract}
%-----------------------------------------------------------------------------%

We present \HI\ spectral line and optical broadband images of the
nearby low surface brightness dwarf galaxy KDG\,215.  The HI images,
acquired with the {Karl G. Jansky Very Large Array (VLA}\footnote{The
  National Radio Astronomy Observatory is a facility of the National
  Science Foundation operated under cooperative agreement by
  Associated Universities, Inc.}), reveal a dispersion dominated ISM
with only weak signatures of coherent rotation.  The HI gas reaches a
peak mass surface density of 6 \msun\,pc$^{-2}$ at the location of the
peak surface brightness in the optical and the UV.  Although KDG\,215
is gas-rich, the H$\alpha$ non-detection implies a very low current
massive star formation rate.  In order to investigate the recent
evolution of this system, we have derived the recent and lifetime star
formation histories from archival Hubble Space Telescope images. The
recent star formation history shows a peak star formation rate $\sim$1
Gyr ago, followed by a decreasing star formation rate to the present
day quiescent state.  The cumulative star formation history indicates
that a significant fraction of the stellar mass assembly in KDG\,215
has occurred within the last 1.25 Gyr.  KDG\,215 is one of only a few
known galaxies which demonstrates such a delayed star formation
history. While the ancient stellar population (predominantly red
giants) is prominent, the look-back time by which 50\% of the mass of
all stars ever formed had been created is among the youngest of any
known galaxy.

\end{abstract}	

\keywords{galaxies: evolution --- galaxies: dwarf --- galaxies:
irregular --- galaxies: individual (KDG\,215)}

%-----------------------------------------------------------------------------%
\section{Introduction}
\label{S1}
%-----------------------------------------------------------------------------%

The interplay between the gaseous and stellar components of low mass
galaxies is extremely complex.  Gas-rich dwarf irregular galaxies
(i.e., systems with significant mass reservoirs of neutral hydrogen,
HI) often host ongoing massive star formation (SF, as traced by
\halpha\ emission, with characteristic timescales of $<$10 Myr) and
have significant ultraviolet (UV) emission (tracing SF on longer
timescales of 100-200 Myr).  The stochastic sampling of the upper
portion of the initial mass function is known to be a complicating
factor in low mass galaxies \citep{lee09}.  While significant progress
has been made in understanding how the properties of the resolved
stellar populations and the integrated UV luminosities are related
\citep{mcquinn15}, it remains very difficult to predict how a certain
gas-rich dwarf irregular galaxy converts its available reservoir of
gas into stars.

\begin{figure*}
\epsscale{1.15}
\plotone{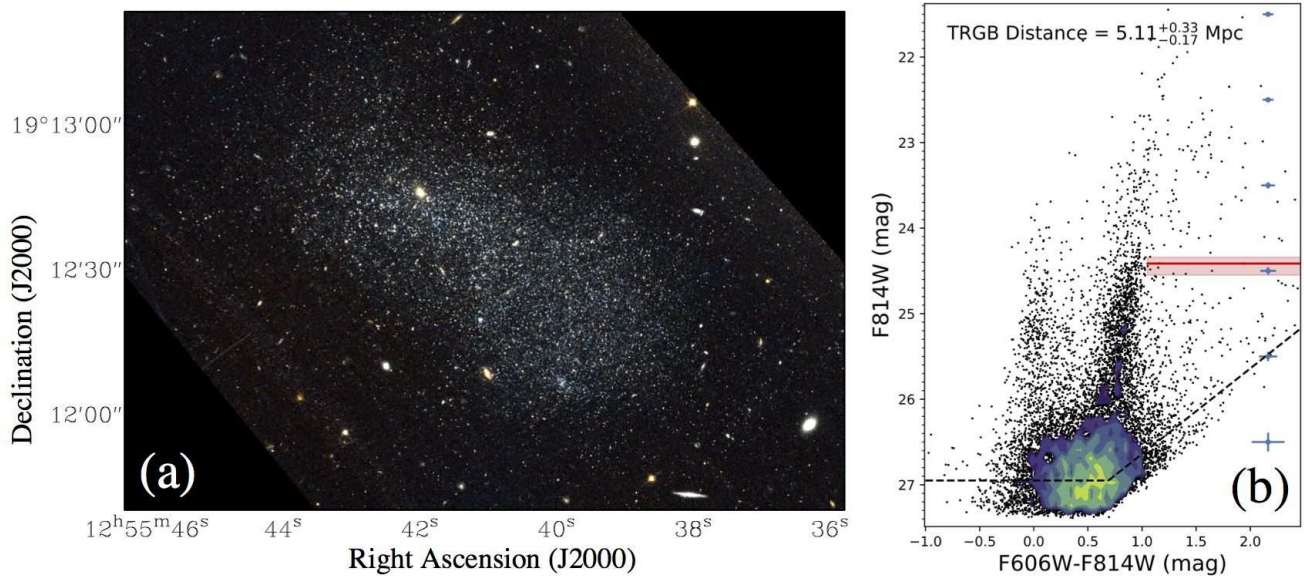}
\caption{HST color image (a) and CMD (b) of KDG\,215.  The CMD reveals
  a strong red giant branch, a significant population of red
  supergiants, a weak blue plume, and red clump stars. The more
  densely populated fainter magnitudes are shown as number density
  contours for clarity.  The TRGB is marked by a red line and shaded
  red area resulting in a distance D $=$ 5.11$^{+0.33}_{-0.17}$ Mpc.}
\label{CMD} 
\end{figure*}

An especially important type of galaxy in this regard is the low
surface brightness dwarf.  Traditionally defined as having low central
surface brightnesses ($\mu_{\rm B}$\,$>$\,23.0 mag arcsecond$^{-2}$),
low surface brightness galaxies span all galaxy masses and
morphological types \citep[e.g.,][]{mcgaugh94,deblok95}.  Many (but
not all) low surface brightness galaxies host ongoing massive SF as
traced by \halpha\ emission \citep[e.g.,][]{schombert11}.  The dwarf
members of this class have been extensively studied in the context of
the threshold gas mass surface density required for SF
\citep[e.g.,][]{vanderhulst93,vanzee97} as well as in the derivation
of high resolution rotation curves and the resulting constraints on
the dark matter distribution within galaxies
\citep[e.g.,][]{deblok01,deblok02}.

Nearby low surface brightness dwarf galaxies offer a unique
opportunity to study their recent SF on a spatially resolved basis.
This allows for the reconstruction of the recent evolutionary pathways
that have led to their current low surface brightness state.  Gas-rich
low surface brightness systems that are quiescent in terms of current
massive SF (i.e., are \halpha\ non-detections at meaningful
sensitivity levels) are especially interesting, as they offer a
glimpse of the conditions in which SF has ceased altogether.  Is this
``quenching'' of SF due to the current conditions of the gas?  Is it
caused by feedback from previous SF?  Is the efficiency of the SF
process fundamentally different than in more massive systems?
Answering these questions provides important empirical constraints on
simulations of the evolution of low-mass galaxies
\citep[e.g.,][]{hopkins14,onorbe15}.

The subject of this work, KDG\,215 (originally cataloged in
{Karachentseva 1968}\nocite{karachentseva68}, also known as
LEDA\,44055, F575$-$3 from {Schombert
  \etal\ 1992}\nocite{schombert92}, or D575$-$5 from {Schombert
  \etal\ 1997}\nocite{schombert97}), is a galaxy that possesses unique
characteristics.  First, its optical surface brightness is extremely
low.  Of the more than 175 irregular galaxies studied in
\citet{hunter06}, KDG\,215 has the second lowest central surface
brightness ($\mu_{\rm V}$\,$=$\,24.69$\pm$0.15 mag arcsecond$^{-2}$).
\citet{schombert11} finds a somewhat higher central surface brightness
($\mu_{\rm V}$\,$=$\,23.80 mag arcsecond$^{-2}$).  Second, KDG\,215
has a current star formation rate (SFR) of zero.  Of the more than 60 low
surface brightness galaxies in \citet{schombert11}, KDG\,215 is one of
only four \halpha\ non-detections (see also {Karachentsev \& Kaisina
  2013}\nocite{karachentsev13}).  Third, the source is nearby and has
sufficiently deep HST images to allow precision color magnitude
diagram (CMD) work.  The distance measurement by
\citet{karachentsev14} places the object securely in the Local Volume
(D $=$ 4.83$\pm$0.34 Mpc).  Fourth, the source is gas rich.  The HI
properties were first measured in \citet{salzer90} and \citet{eder00},
in which the total HI flux integrals were measured to be S$_{\rm HI}$
$=$ 4.48 Jy\,km\,s$^{-1}$ and 4.37 Jy\,km\,s$^{-1}$, respectively.
The recently completed ALFALFA catalog \citep{haynes18} revises the
total HI flux integral up to S$_{\rm HI}$ $=$ 5.51\,$\pm$\,0.06
Jy\,km\,s$^{-1}$.  Finally and most importantly, as we demonstrate in
this manuscript, KDG\,215 has a star formation history (SFH) that is
extreme compared to that of any other known dwarf galaxy: a
significant fraction of the stellar mass has been formed within the
last 1.25 Gyr.

%-----------------------------------------------------------------------------%
\section{Observations and Data Handling}
\label{S2}
\subsection{VLA HI Observations}
\label{S2.1}
%-----------------------------------------------------------------------------%

New HI spectral line observations of KDG\,215 were acquired with the
Karl G. Jansky Very Large Array (VLA) in the C configuration on July
31, 2017 for program TDEM0025.  This investigation was made possible
by the National Radio Astronomy Observatory's ``Observing for
University Classes'' program.  Details about the program and its uses
can be found in \citet{cannon17}.

\begin{figure*}
\epsscale{1.15}
\plotone{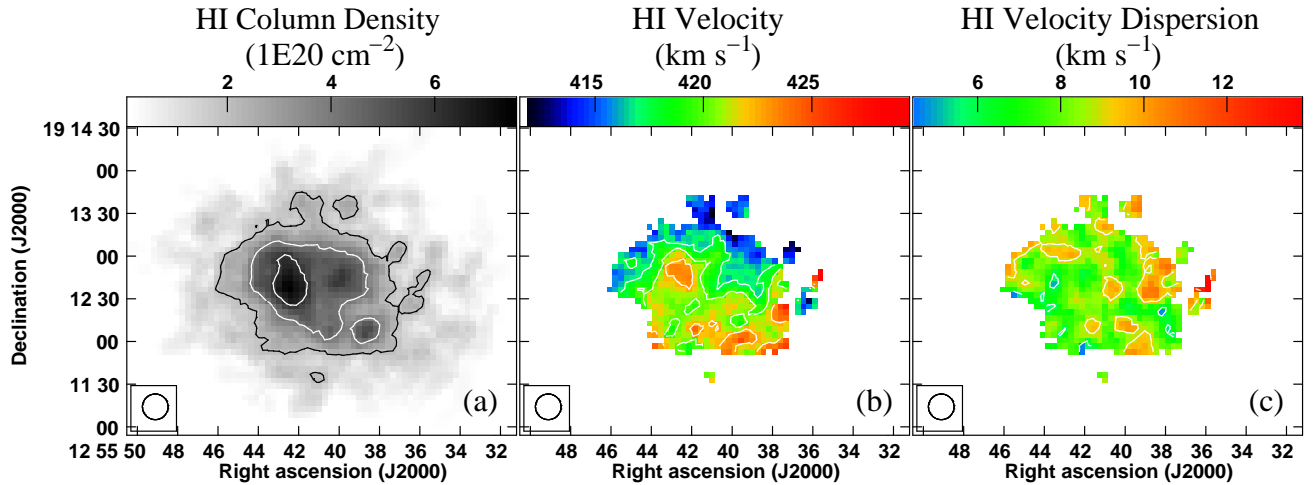}
\caption{HI images of KDG\,215.  Panel (a) shows the HI column density
  distribution in units of 10$^{20}$ cm$^{-2}$; contours are at levels
  of (2, 4, 6)\,$\times$\,10$^{20}$ cm$^{-2}$.  Panel (b) shows the
  intensity weighted velocity field in units of km\,s$^{-1}$; contours
  are at levels of 417.5, 420, and 422.5 km\,s$^{-1}$.  Panel (c)
  shows the intensity weighted velocity dispersion; contours are at
  levels of 6, 9, and 12 km\,s$^{-1}$.  Panels (b) and (c) are
  threshold blanked using the column density image shown in panel (a)
  with a threshold of 2\,$\times$\,10$^{20}$ cm$^{-2}$ (i.e., the
  outermost contour shown in panel a).  The circular beam size of
  18\arcsec\ (shown in the bottom left of each panel) corresponds to a
  physical resolution of 446 pc at the adopted distance of 5.11 Mpc.}
\label{KDG215HI} 
\end{figure*}

Observations of KDG\,215 [$\alpha$,$\delta$ (J2000) $=$ 12$^{\rm h}$
  55$^{\rm m}$ 41.$^{\rm s}$0, $+$19\arcdeg 12\arcmin 33\arcsec] were
acquired in a two-hour observing block that used 3C286 as the primary
calibrator and J1330$+$2509 as the secondary calibrator.  The total on
source integration time for KDG\,215 was approximately 91.6 minutes.
The spectral window centered on the HI spectral line was 16 MHz wide
and divided into 4096 channels, delivering a native velocity
resolution of 3.906 kHz\,channel$^{-1}$ (0.82
km\,s$^{-1}$\,channel$^{-1}$).  The data were calibrated using
standard formalisms in the CASA\footnote{https://casa.nrao.edu}
environment.  Continuum subtraction was performed using a first order
fit to line-free channels.

Imaging of the continuum subtracted visibilities was performed in a
multi-step procedure using the CLEAN task in CASA.  A Gaussian $uv$
taper length of 10 k$\lambda$ was employed when using the Briggs
weighting scheme with ROBUST$=$0.5.  The velocity resolution was set
to 2.5 km\,s$^{-1}$ (i.e., $\sim$3 times larger than the native
spectral resolution).  The original synthesized beam
(16.11\arcsec\,$\times$15.46\arcsec) was smoothed spatially to a
circular 18\arcsec.  Cleaning was performed to the 1$\sigma$ level
(1.6 mJy).

Moment maps were created by threshold blanking at the 1$\sigma$ level
and then blanking by hand to ensure velocity coherence.  The moment
zero map produced in this manner yields a total HI flux integral
S$_{\rm HI}$ $=$ 4.84\,$\pm$\,0.48 Jy km\,s$^{-1}$.  This image was
calibrated into units of column density (specifically, 10$^{20}$
cm$^{-2}$) and used as a transfer mask against the moment one and
moment two maps at the level of N$_{\rm HI}$ $=$
2\,$\times$\,10$^{20}$ cm$^{-2}$.  The total HI flux integral found in
the moment zero map discussed above is slightly smaller than the
ALFALFA total flux integral (5.51\,$\pm$\,0.06 Jy\,km\,s$^{-1}$;
{Haynes \etal\ 2018}\nocite{haynes18}).  To optimize surface
brightness sensitivity, a low resolution (Gaussian $uv$ taper length 4
k$\lambda$), single channel image was created.  This image is centered
at the ALFALFA systemic velocity (V$_{\rm HI}$ $=$ 419 \kms) and has a
channel width equal to the ALFALFA W$_{\rm 50}$ (22\,$\pm$\,2
\kms). This technique places all of the HI line emission into a single
channel and samples only velocities with significant HI flux.  The
resulting HI moment zero image recovers S$_{\rm HI}$ $=$
5.50\,$\pm$\,0.55 Jy km\,s$^{-1}$, in excellent agreement with the
ALFALFA HI flux integral.

%-----------------------------------------------------------------------------%
\subsection{Hubble Space Telescope Observations}
\label{S2.2}
%-----------------------------------------------------------------------------%

KDG\,215 was observed with the Advanced Camera for Surveys (ACS)
onboard the Hubble Space Telescope (HST) on 2012 November
20 for program GO-12878 (P.I. Karachentsev).  The F606W and F814W
filters were used.  Integration times of 2076 sec (F814W) and 1640 sec
(F606W) were achieved in two separate exposures per filter.  Standard
processing in the ACS pipeline was applied, including corrections for
charge transfer efficiency.

Single-star photometry was performed on the HST images using the ACS
module of the DOLPHOT package \citep{dolphin00}.  Well-measured stars
were selected by implementing cuts based on sharpness and crowding.  A
minimum S/N ratio of 5 is required for a star to be included in the
final photometry lists, which were then corrected for foreground
extinction (A$_{\rm F606W}$ $=$ 0.049 mag; A$_{\rm F814W}$ $=$ 0.030
mag) via the \citet{schlafly11} recalibration of the
\citet{schlegel98} dust maps. Artificial star tests were used to
determine the 50\% completeness limits of the data (27.69 mag in F606W,
26.98 mag in F814W).

%-----------------------------------------------------------------------------%
\section{The Resolved Stellar Population of KDG\,215}
\label{S3}
%-----------------------------------------------------------------------------%

Figure~\ref{CMD}(a) shows the color HST image of KDG\,215 (F606W as
blue, F814W as red, and the linear average of the two filters as
green).  The low surface brightness nature of the stellar population
of KDG\,215 is readily apparent.  The stellar population has a total 
physical extent of $\sim$2 kpc, and is dominated by the emission from 
stars with blue colors.  

The F814W vs. (F606W$-$F814W) CMD is shown in panel (b) of
Figure~\ref{CMD}.  The CMD is populated with more than 11,000
individual stars, distributed in four main regions: the blue plume
(containing main sequence and blue helium burning stars), the red
giant branch, the red supergiant region, and red clump stars.  The
density of stars at fainter magnitudes is higher, and so these regions
are plotted as number density contours for ease of interpretation.

\begin{figure*}
\epsscale{1.15}
\plotone{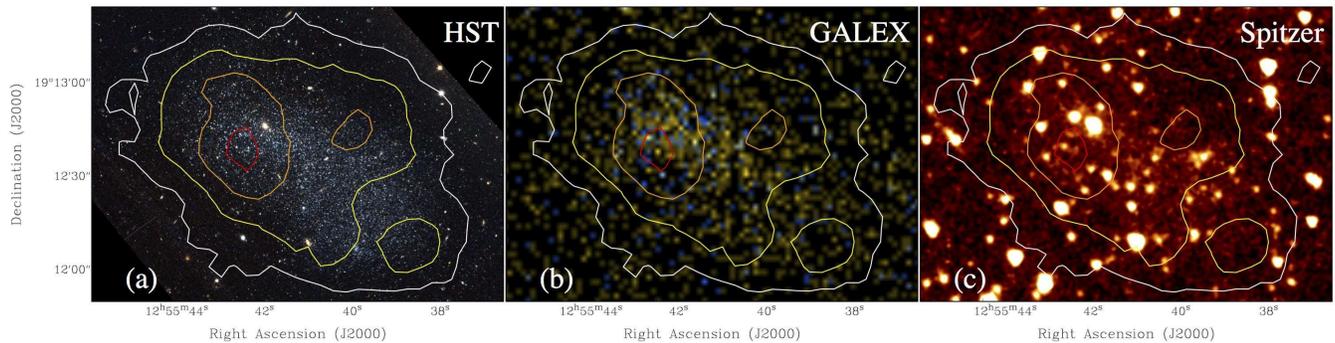}
\caption{Comparison of HI surface density and the stellar populations
  of KDG\,215.  Panel (a) shows a color HST image (F606W in blue,
  F814W in red, and a linear average of the two as green); panel (b)
  shows a color GALEX image (FUV in blue, NUV in red, and a linear
  average of the two as green); panel (c) shows a Spitzer 3.6\,$\mu$m
  image.  Contours of HI column density (18\arcsec\ beam; same image
  as shown in Figure~\ref{KDG215HI}) are overlaid at levels of
  2.5\,$\times$\,10$^{20}$ cm$^{-2}$ (white), 4\,$\times$\,10$^{20}$
  cm$^{-2}$ (yellow), 5.5\,$\times$\,10$^{20}$ cm$^{-2}$ (orange), and
  7\,$\times$\,10$^{20}$ cm$^{-2}$ (red).}
\label{KDG215HIstellar} 
\end{figure*}

From this CMD, the TRGB distance is identified as
5.11$^{+0.33}_{-0.17}$ Mpc (shown by the thick red line and shaded
region of the CMD).  This value is slightly larger than the previous
TRGB estimate in \citet{karachentsev14}, although the measurements
agree within errors.  At this distance, the HI mass of KDG\,215 is
M$_{\rm HI}$ $=$ (3.40$\pm$\,0.34)\,$\times$\,10$^7$ \msun.  The
current stellar mass (using the total stellar mass formed and a return
fraction of 0.412 from {Vincenzo \etal\ 2016}\nocite{vincenzo16}; see
discussion in \S~\ref{S5}) is M$_{\star}$ $=$
(8.0$^{+1.2}_{-3.2}$)\,$\times$\,10$^{6}$ \msun.

%-----------------------------------------------------------------------------%
\section{The Neutral Gas Morphology and Dynamics of KDG\,215}
\label{S4}
%-----------------------------------------------------------------------------%

HI gas is detected at the 3$\sigma$ level or above over $\sim$35
\kms\ in the final data cube.  The collapse of the 3-dimensional
datacube into 2-dimensional moment maps produces the images presented
in Figure~\ref{KDG215HI}.  Panel (a) shows the HI mass surface density
image in units of 10$^{20}$ cm$^{-2}$.  The HI gas is well-resolved by
the beam (roughly 10 full beam widths across the galaxy).  The peak HI
column density is 7.5\,$\times$\,10$^{20}$ cm$^{-2}$, which
corresponds to an HI mass surface density N$_{\rm HI}$ $=$ 6
\msun\,pc$^{-2}$.  Figure~\ref{KDG215HIstellar} compares the HI column
densities (shown as contours) with images from HST (analyzed in detail
below), GALEX (tracing SF on 100-200 Myr timescales), and Spitzer
(tracing the majority of the stellar mass, though significantly
contaminated by foreground and background objects).  The HI mass
surface density maximum is roughly co-spatial with the region with the
highest UV and optical flux.

Figures~\ref{KDG215HI}~(b) and (c) show the intensity weighted
velocity field and velocity dispersion, respectively.  The HI
kinematics of KDG\,215 are complex.  There is a weak signature of
projected rotation with magnitude $\sim$10 \kms, oriented more or less
south-to-north.  However, there are no single values for the kinematic
major or minor axes.  Further, the magnitude of the projected rotation
($\sim$10 \kms) is only slightly larger than the average HI velocity
dispersion throughout the disk ($\sim$8-10 \kms).  As discussed in
detail in \citet{mcnichols16}, this signifies the empirical transition
between galaxies with obvious rotational support (where rotation
velocity largely exceeds the HI velocity dispersion) and galaxies
which are pressure supported.  Attempts to fit tilted ring models to
the HI velocity field were not successful.  Fitting the 3-dimensional 
cube with Gaussians to extract a velocity field  produced 
no significant changes from the intensity weighted image shown in 
Figure~\ref{KDG215HI}~(b).  With the current HI data we are unable to
measure the dynamical mass of KDG\,215 with confidence.

%-----------------------------------------------------------------------------%
\section{The Star Formation History of KDG\,215}
\label{S5}
%-----------------------------------------------------------------------------%

\begin{figure*}
\epsscale{1.15}
\plotone{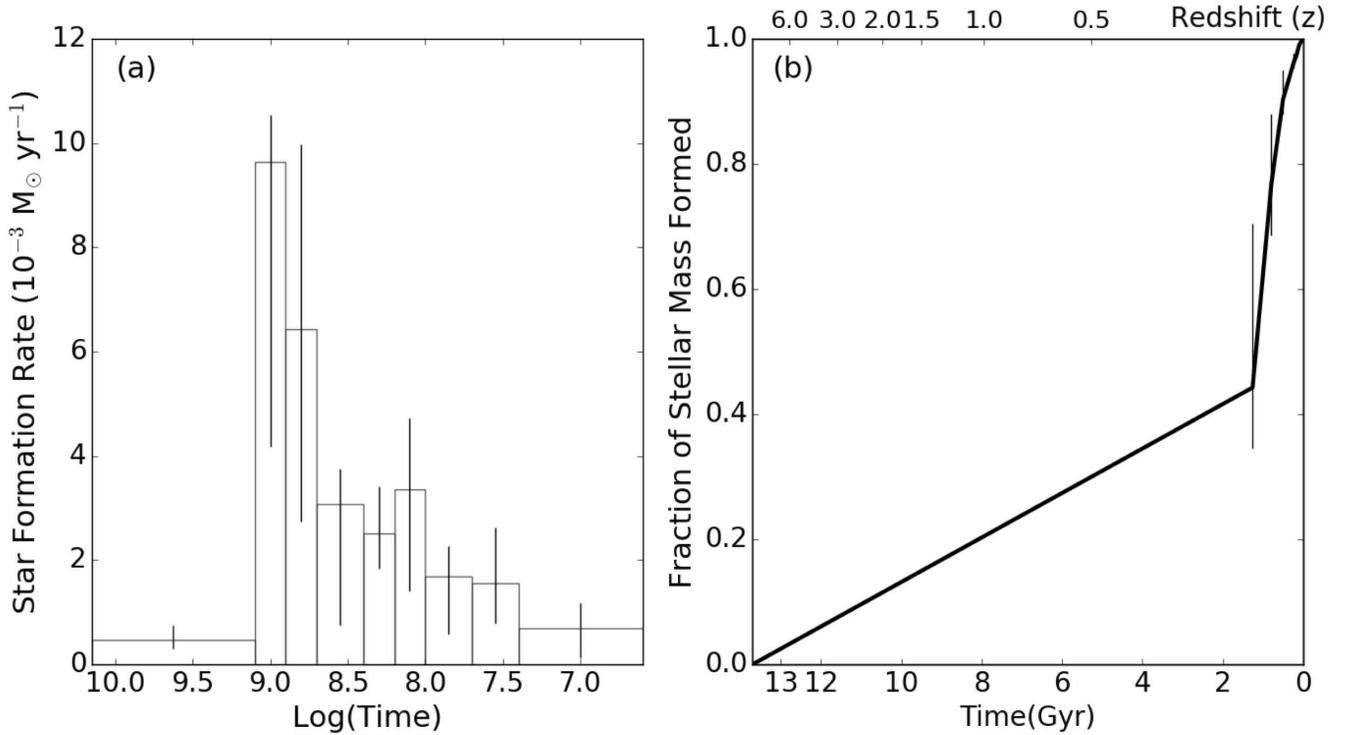}
\caption{The SFH of KDG\,215.  Panel (a) shows the SFR as a function
  of time in the galaxy, while panel (b) shows the cumulative SF as a
  function of time in the galaxy.  The SF event that peaked in
  intensity $\sim$1 Gyr ago formed more than 50\% of the total stellar
  mass in KDG\,215. The median look-back time by which 50\% of the
  mass of all stars ever formed had been created in KDG 215 is of the
  order a Gyr.}
\label{HST_SFH} 
\end{figure*}

Figure~\ref{HST_SFH} presents the SFH of KDG\,215 as derived from the
HST images.  These plots are created using the CMD-fitting algorithm
{\sc{MATCH}} \citep{dolphin02}, to which we refer the reader for
details.  Very simply, {\sc{MATCH}} creates simple stellar populations
that are combined to create a best-fit CMD based on a Poisson
likelihood statistic.  The photometric depth of our data reaches
$\sim$2 mag below the tip of the red giant branch (see
Figure~\ref{CMD}), allowing us to explore the most recent $\sim$1 Gyr
with significant time resolution.  The HST data do not have sufficient
photometric depth to parse the SF episodes beyond $\sim$1 Gyr, and so
everything older than 1.25 Gyr is grouped into one single time bin.
For details on the temporal resolution of SFHs derived from HST data,
see the discussion in \citet{mcquinn10a}.

Figure~\ref{HST_SFH}(a) shows the global SFR of KDG\,215 as a function
of (logarithmic) time.  This demonstrates that KDG\,215 had a low SFR
for most of the Hubble time.  Within the most recent Gyr, KDG\,215
underwent a significant episode of strong SF.  The peak intensity
($\sim$0.01 M$_{\odot}$\,yr$^{-1}$) occurred around 1 Gyr ago, and
then fell steadily to the current rate (consistent with zero).  This
SF event produced more than half of the total stellar mass formed in
KDG\,215 (see discussion below).

Figure~\ref{HST_SFH}(b) shows the cumulative SFH of KDG\,215 as a
function of (linear) time.  The fraction of the total stellar mass
ever formed in the galaxy [M$_{\star}$ $=$
  (1.38$^{+0.20}_{-0.56}$)\,$\times$\,10$^{7}$ \msun] rises from zero
to unity at $z$ $=$ 0.  The first time bin is large, covering log(age)
9.1-10.15 (corresponding to look-back times 1.25 Gyr - 14.1 Gyr).  At
the end of this first time bin, the fraction of the total stellar mass
formed is 0.44$^{+0.26}_{-0.10}$.  Interpreted at face value, the
galaxy had formed only 44\% of its total stellar mass.  Due to the
limitations imposed by the photometric depth, the uncertainties on
this fraction (which include both random and systematic terms, with
the latter dominating the error budget) allow for a range between 34\%
and 70\%.  From the best-fitting SFH, the median look-back time by
which 50\% of the mass of all stars ever formed had been created in
KDG\,215 is of the order a Gyr.  It is important to note that while
the median age of the stellar population is young, ancient stellar
populations (predominantly red giants; see Figure~\ref{CMD}) are
present.

%-----------------------------------------------------------------------------%
\section{Interpretation}
\label{S6}
%-----------------------------------------------------------------------------%

The delayed stellar mass assembly of KDG\,215 makes it an important
test of models of structure formation.  The significant burst of SF
that occurred $\sim$1 Gyr ago is qualitatively similar in terms of
relative strength and duration to the strongest bursts seen in the
simulations of \citet{governato15}.  The primary difference is that
this SF episode occurs at a much later epoch in KDG\,215 than those
seen in the simulations.  Baryonic physics could allow for this
delayed stellar mass assembly.  As discussed in \citet{christensen16},
the efficiency of outflows increases as galaxy mass decreases.  If
early SF (i.e., activity at $z$\,$\gsim$\,0.5, in the oldest time bin
shown in Figure~\ref{HST_SFH}) drove significant outflow episodes that
were followed by eventual gas recycling, then the overall SFR in
KDG\,215 would be low over cosmological timescales.

\begin{figure*}
\epsscale{0.65}
\plotone{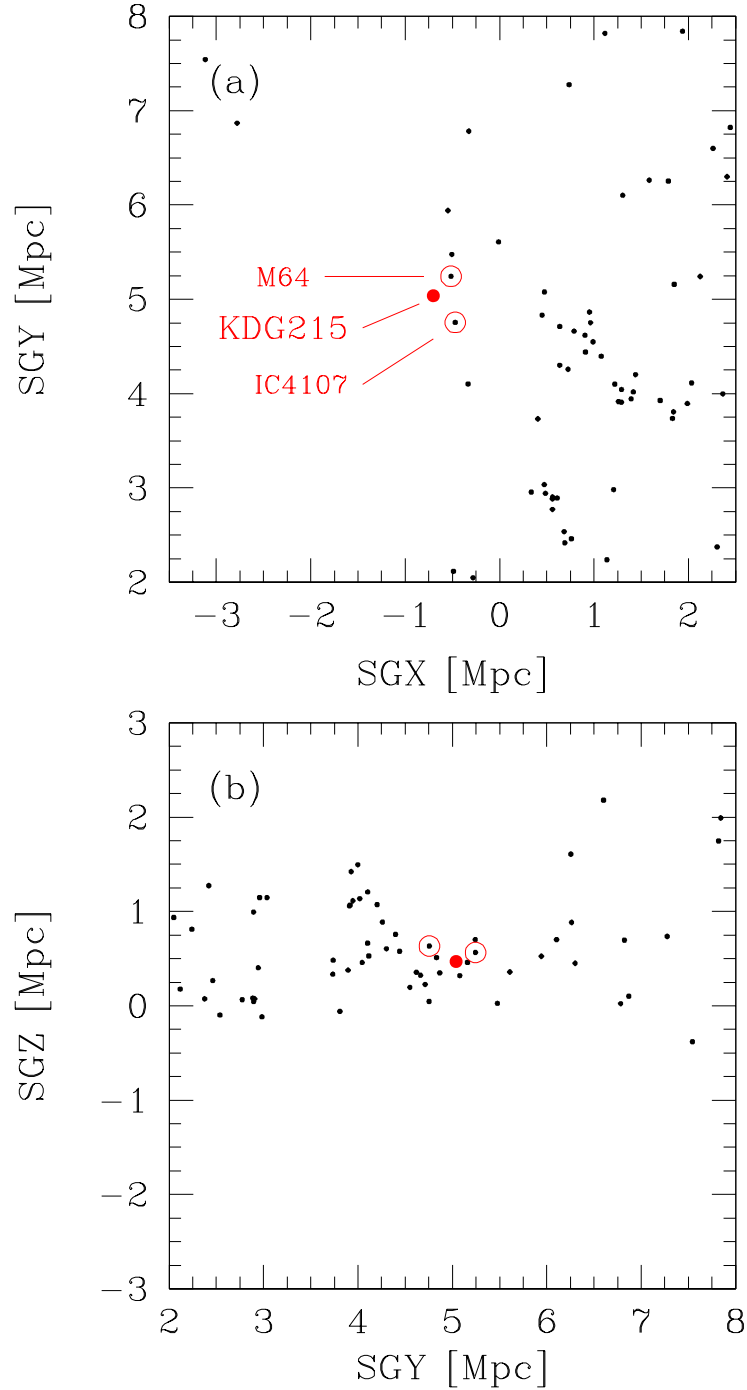}
\caption{The nearby environment of KDG\,215.  All galaxies from the
\citet{tully16} database that are located within a cubical volume 6
Mpc on a side and roughly centered on KDG\,215 are plotted.  The
nearest neighbor of KDG\,215 is the spiral galaxy M\,64; the two
galaxies have a physical separation of 295 kpc.  Also labeled is
IC\,4107 \citep[physical separation of $\sim$400 kpc;][]{karachentsev18}.}
\label{KDG215_LSS} 
\end{figure*}

The triggering mechanism of the starburst episode $\sim$1 Gyr ago may
be related to the local environment of KDG\,215.
Figure~\ref{KDG215_LSS} plots the position of KDG\,215 along with all
galaxies from the \citet{tully16} database located in a 6 Mpc cube
centered roughly on KDG\,215.  Positions are plotted in Supergalactic
coordinates.  Panel (a) shows the SGX vs. SGY plane (i.e., looking
down on the Supergalactic plane), while panel (b) shows the SGY
vs. SGZ plane (in which the Supergalactic plane is clearly defined).
The nearest neighbor galaxies of KDG\,215 are the spiral M\,64 and the
irregular galaxy IC\,4107 (also known as KK\,177; {Karachentseva \&
  Karachentsev 1998}\nocite{karachentseva98}).  For KDG\,215 at 5.11
Mpc (this work), M\,64 at 5.30 Mpc \citep{tully16}, and KK\,177 at
4.82 Mpc \citep{karachentsev18}, the physical separations are 295 kpc
(KDG\,215 and M\,64) and $\sim$400 kpc (KDG\,215 and KK\,177).

If KDG\,215 encountered the higher density environment of M\,64 a few
Gyr ago, then this may have initiated the SF event.  This scenario is
similar to the reignition of SF seen in simulations of low-mass field
dwarfs \citep{wright18}. SF is quelled by stellar feedback or
reionization but much of the gas heated by these mechanisms remains in
the halos of the galaxies. Interactions with low density streams of
gas in the local environment of the galaxies compresses the gas in the
halos inducing SF. While the characteristics of such suppression and
reignition is typically only discernable in SFHs with higher temporal
resolution (created from a CMD that reaches below the old main
sequence turn-off), the period of quiescence in KDG\,215 is so
extended that the suppression and reignition is apparent in
Figure~\ref{HST_SFH}.

KDG\,215 is one of only a few known galaxies which demonstrates such a
delayed SFH. The closest known analogs are the Local Group galaxies
Leo\,A \citep{cole07} and the Aquarius dIrr \citep{cole14}.
Like KDG\,215, both systems have a young median look-back time for
forming 50\% of their stellar mass (4.2 Gyr for Leo A and 6.8 Gyr for
Aquarius) and low SFRs.  Interestingly, KDG\,215 has larger HI and
stellar masses than both Leo\,A (M$_{\star}$ $=$
3.3\,$\times$\,10$^{6}$ \msun, M$_{\rm HI}$ $=$
1.1\,$\times$\,10$^{7}$ \msun; {Kirby \etal\ 2017}\nocite{kirby17},
{Cole \etal\ 2007}\nocite{cole07}) and Aquarius (M$_{\star}$ $=$
(1-2)\,$\times$\,10$^{6}$ \msun, M$_{\rm HI}$ $=$
2.7\,$\times$\,10$^{6}$ \msun; {Cole \etal\ 2014}\nocite{cole14}).
Further, it is slightly more gas-rich (M$_{\rm HI}$/M$_{\star}$) than
both systems.

The distance of KDG\,215 does not allow us to parse the ancient SFH
with the temporal precision that was achieved for Leo\,A and Aquarius.
The full error budget on the fraction of total stellar mass formed at
a look-back time of 1.25 Gyr implies that KDG\,215 may have formed as
much as 70\% or as little as 34\% of its total stellar mass at this
point.  However, it is important to stress that even if the fraction
is 70\%, KDG\,215 still stands out as extreme in comparison to Leo\,A
and Aquarius.  At a look-back time of 1.25 Gyr, both Leo\,A and
Aquarius have each formed more than 90\% of their total stellar mass
(see Figure~8 of {Cole \etal\ 2014}\nocite{cole14}).  The SFH of 
KDG\,215 is both unique and extreme.

%-----------------------------------------------------------------------------%
\acknowledgements
%-----------------------------------------------------------------------------%

The authors thank the NRAO for making the ``Observing for University
Classes'' program available to the astronomical community.
I.K. acknowledges support by RFBR grant 18-02-00005.

\textit{Facilities}: \facility{VLA}, \facility{HST}

%-----------------------------------------------------------------------------%
\clearpage
\bibliographystyle{apj}                                                 

%-----------------------------------------------------------------------------%

\end{document}